# ON THE QUESTION OF WAVEFUNCTION COLLAPSE IN A DOUBLE-SLIT DIFFRACTION EXPERIMENT

FOTINI PALLIKARI

*Faculty of Physics, Department of Solid State Physics, Panepistimiopolis Zografou, Athens 157 84, Greece*

**Abstract**. It was recently proposed that the fringe visibility in the interference pattern of a double-slit diffraction experiment with a HeNe laser beam was reduced as a consequence of the focused attention of participants on the double-slit that initiated the collapse of the photon wavefunction. A thorough examination of the reported apparatus and diffraction pattern data is presented here revealing a number of inconsistencies in the experimental setup. The hypothesis for wavefunction collapse under the proposed experimental conditions is found problematic as the use of single photon sources is demanded.

## 1 INTRODUCTION

Bohr's principle of complementarity of the wave-particle duality is demonstrated in the double-slit interference experiment. When the path of a particle through each of the two slits is determined, its particlelike property is expressed. On the other hand, the wavelike property is documented by the clearly visible interference fringes. The observation of interference pattern and the acquisition of which-path information of the path of the particle through one slit are, therefore, mutually exclusive in this so-called interferometric complementarity [1].

The degree of wave interference or contrast between fringes, i.e. the wave-like quality of the particle, is measured by the visibility parameter V, estimated through $I_{max}$ & $I_{min}$, the maximum and minimum intensities in the interference pattern, [2]

$$V = \frac{I_{max} - I_{min}}{I_{max} + I_{min}} \qquad (1)$$

Visibility equal to 1 corresponds to the highest contrast between fringes, a sign of wavelike quality, when $I_{max}$ and $I_{min}$ take their highest and lowest values, respectively. On the other hand, the which-path information parameter (WPI) or the distinguishability parameter P, quantifies the available paths to a particle. If the particle is equally likely to take any of the available paths then P is 0 depicting again its wavelike quality. If a particular path is much more likely than any other path then P is close to 1. The particlelike property is therefore witnessed as the path of the particle is identified. In the conventional understanding of quantum mechanics we have no right to claim that any given particle actually follows a definite path, unless we actually measure the path of each one of them [3]. It is possible, however, to have partial knowledge of the particlelike as well as the wavelike property simultaneously.

Partial which-path information and reduced visibility can be obtained as long as the quantities P and V fall within the complementarity inequality [4, 5]

$$P^2 + V^2 \leq 1 \qquad (2)$$

This duality relation puts an upper bound to the maximum values of simultaneously determined interference visibility V and path distinguishability P. The equality in (2) holds when each of the two beams passing through the slits are perfectly coherent. Relation (2) has been experimentally confirmed in a variety of situations [6, 7]. In experiments involving photons it is demanded that a single photon source is used so that full and unambiguous WPI is obtained to justify the particlelike property observed, complementary to the observation of interference [8, 9, 10]. The studied phenomena are recognized as quantum phenomena provided the single photon condition is fulfilled.

The probability to find the photon at a time and place is determined by its wavefunction. When there is enough information to identify its path, this is said to cause the collapse of its wavefunction, since what was before just a set of probabilities has now been replaced by a certainty regarding the photon's trajectory. Without the prerequisite of single photons, the study of any trends observed in a double-slit diffraction experiment is simply treated by classical electrodynamics rather than quantum theory.

To measure the visibility parameter (V) in an interference pattern is a relatively straightforward task. The estimation of the distinguishability parameter (P) of the path of a photon, on the other hand, calls for a special experimental setup as, for instance, one that enables the entanglement between the single photon's path and its polarization by introducing an adjustable half wave plate [6], or by the introduction of a second quantum system to serve as a which-path detector [8].

There were also attempts to distinguish the path of photons by unconventional methods, such as by employing the process of extra sensory perception. The more recent of these experiments has used the beam of a HeNe laser diffracted through a double-slit to investigate the role of consciousness in the collapse of the wavefunction (reference [11] and related references therein). The working hypothesis in ref. [11] suggests a decrease in the contrast of interference fringes due to the wavefunction collapse of the photons. It therefore suggests a decrease in the visibility (V) parameter, as a consequence of the increase in the distinguishability (P) of the path of the photons accomplished by mentally focusing on the invisible photon path. To investigate any changes in fringe visibility, the intensity of light in the interference pattern was recorded on a line of 3000 pixels of a digital camera. These digital recordings fed to a computer were further subjected to a fast Fourier transform analysis (FFT) from which the power spectrum was extracted and plotted as power against frequency expressed in wavenumbers. The power $P_K$ corresponding to the interference pattern appearing as a peak at wavenumber K was divided by the power $P_1$ at wavenumber 1 to yield the ratio R. The R parameter was predicted to decrease by the focusing of conscious attention on the double-slit. It is shown in appendix I that the R parameter, the experimental measure to be tested in ref. [11], is related to the visibility V of the double-slit diffraction pattern.

However, the publication of Radin et al does not report absolute values of R during the course of an experiment (or absolute values of the fringe visibility), but the end result of a normalization yielding the parameter $R_Z$, instead. At the end of each session the average ($\mu$) of all R values was estimated, as well as their standard deviation from the mean ($\sigma$). The normalized $R_Z$ parameter was obtained by subtracting $\mu$ from each R and dividing their difference by $\sigma$. To observe the role of consciousness in the purported collapse of the photon wavefunction, the power spectrum was recorded under two consciousness-related conditions namely, attention focused toward the double-slit apparatus during the diffraction experiment as compared to away from it. The "attention-toward" condition claimed to allow for partial observation of the photon path by a process not yet scientifically identified. A comparison of the visibility-related parameter R between the two conditions, "attention-toward" and "attention-away", was carried out and the statistical significance of their difference was estimated. The article reports that the statistical significance of this difference is beyond chance expectation and considers it as evidence that focused attention collapses the photon wavefunction. The experiment investigates in addition whether the reported focused attention is indeed taking place during the "attention-toward" experimental condition, (when the fringe visibility is being reduced), by studying an electrocortical marker of attention in electroencephalograph (EEG) spectra. The article reports positively correlated changes between the reduction of fringe visibility and the shifts in attention, confirming that focused attention is indeed taking place during the "attention-toward" condition.

In the next section a thorough evaluation of the experimental setup of ref. [11] is carried out, before the required conditions for observing a wavefunction collapse in a double-slit diffraction experiment are discussed in section 3.

## 2. CHARACTERISTICS OF THE DOUBLE-SLIT DIFFRACTION PATTERN IN RADIN ET AL.

The double-slit diffraction apparatus used in ref. [11] consists of a 5 mW HeNe laser operating at wavelength $\lambda$=632.8nm. The beam power is attenuated to 0.5mW by use of an appropriate filter before it reaches the set of two slits, each of width a=10 μm and a reported distance d=200 μm between their centers. The camera is thus exposed to an even lower intensity of diffracted laser light. The double-slit is located at a distance D = 10.4 cm from a camera, which records the interference pattern by use of a line of 3000 pixels. The size of each camera pixel is wrongly reported in ref. [11] to be 0.2μm by 7μm. It was apparently copied from a misprint in the camera manual (where the 200μm appears with a dot in front of it as .200μm) and misinterpreted to read 0.2μm. The correct size, 7μm by 200μm, is found at two other places inside the manual which is available at the company's site [12].

On the basis of the above data it is possible to estimate certain characteristic features of the diffraction pattern generated by the apparatus of Radin et al, first checking that the necessary condition for Fraunhofer diffraction is satisfied: $a^2/D\lambda \ll 1$. In particular, the reported data allow for the estimation of the following characteristics (see appendix I): (a) the missing order fringe and the number of fringes inside the principal maximum, (b) the width of the principal maximum, (c) the fringe separation at the plane of the camera, (d) the position of the first secondary maximum and (e) the position of the characteristic double-slit interference power peak in the FFT spectrum. A sketch of the double-slit setup is presented in Fig. 1 for clarity, at the right hand side of which the diffraction pattern reported in ref. [11] is added.

A disagreement between the characteristics of the double-slit diffraction pattern in Radin et al and the reported apparatus description becomes apparent, see appendix I. The principal maximum should be only 1884 pixels wide rather than the reported 2308 pixels in fig. 2A, the fringe separation inside the principal maximum should be 47 pixels instead of the 69 pixels reported, the absent first secondary maximum should be already in view at the 2869th pixel having about 282 arbitrary units height and finally, there should appear 39 fringes inside the principal maximum instead of the observed 32.

These discrepancies, the broader principal maximum, the larger fringe separation and the absence of the first secondary maximum could be due to a couple of reasons. One being a possible larger distance between the camera and the double-slit, i.e. about 14 cm instead of the reported 10.4 cm, see appendix I. They could also be due to the actual irregular shape of each slit, which may differ from the exact rectangular with sharp edges. Rectangular slits may have uneven width, due for instance to the relatively large size of their drilling tool. In such cases, suppression of the secondary maxima occurs, an effect known as "apodization", followed by the broadening of the principal maximum, as the case could have been here. Apodization is, in fact, beneficial for certain applications where the presence of secondary maxima blurs the image of a projected object and techniques are employed to artificially

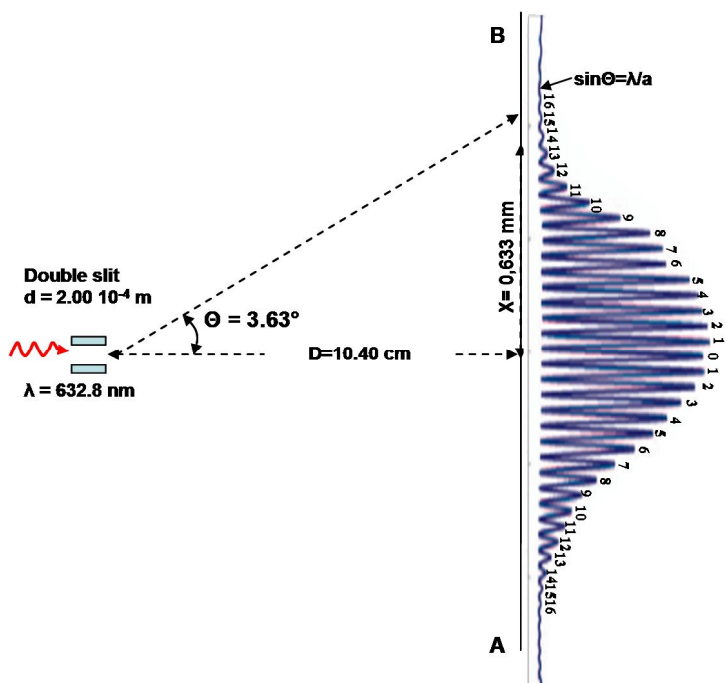

Fig. 1. The double-slit diffraction experimental setup of ref. [11]

introduce it in their optical systems [13-16].

Even though some discrepancies between reported data in Radin et al could be understood on the assumption that either the camera was positioned differently than was reported or, that a not-so-ideally shaped double-slit was used, such grounds cannot account for the 7 missing fringes inside the principal maximum. A possible reason for the absent fringes could be camera pixels that are not sensitive enough because of contamination and therefore not adequately recording the intensity distribution of the diffracted light. An effort to produce the diffraction pattern by a solid line, speculatively drawn on a scatter of deficient pixel recordings, as the case may be in fig. 2A of ref. [11], can miss a few fringes.

The experimental data points should have already been plotted and be present on this diffraction pattern, as one would expect from an analogous scientific experiment [17]. The existence of a solid line that highlights the shape of fringes in a diffraction pattern plotted on absent data points is then highly questionable. Was this line drawn on the basis of fitting the (currently invisible) experimental data points on a mathematical function representing the intensity distribution, in which case what was the form of this mathematical function?

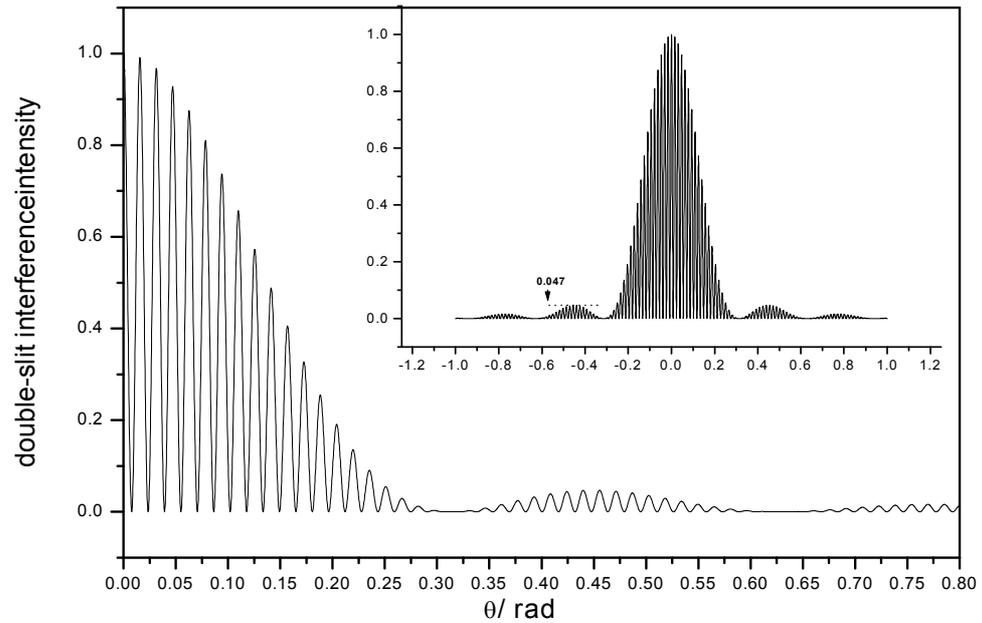

Fig. 2 Simulated double-slit diffraction pattern. The order of the missing fringe is $m_{missing\,order}$=20. See text for details.

A statistical approach that tests a hypothesis comparing results between two conditions does not assign minor importance to questions such as the above. A not sensitive enough line of camera pixels, that is corroborated by the irregular shape of the principal maximum in the diffraction pattern, produces unreliable records of data from one test to the other that are supposed to confirm a hypothesis, i.e. that the fringe visibility is perturbed by conscious attention.

Furthermore, as shown in appendix I-(e) and II, the power spectrum from which the R parameter is estimated, the fundamental factor of analysis in Radin et al, was placed at the wrong position on the graph. It needs to be shifted from its current position to the left by about one wavenumber before the measure R is estimated. The FFT analysis of Radin et al finds the main power peak at 45 wavenumbers, see fig. 2B and its caption. The necessity for shifting the power spectrum is justified also by another fact. In its new position the power peak is located at 44 wavenumbers, which happens to be the inverse (in wavenumbers) of the fringe separation of 69 pixels, as expected (see appendix I). More importantly, the power at 1 wavenumber at the new position will be significantly higher than it is currently displayed in fig. 2B, due to the steepness of the power spectrum at low frequencies. By bringing the spectrum to the corrected position after the shift, the measure R can be estimated from fig. 2B of Radin et al to be R≅0.32 from which a good estimate of the visibility parameter turns out to be V≅75%, see appendix II. On the other hand, the visibility estimated from fig. 2A is close to 100% indicating incompatibility of the two figures in terms of the estimation of V, unlike what would be expected given that the one is derived from the other.

## 3. Discussion

The intensity distribution of the double-slit diffraction pattern in Radin et al deviates from that expected on the basis of the reported setup. Fringes are missing from inside the central maximum. The intensity of the present fringes appears unevenly shaped indicating a combination of causes such as a possible angle of the double-slits to the vertical, or a possible contamination of the camera pixels affecting their sensitivity. The experimental diffraction pattern data points are not presented on the graph, its shape is nevertheless outlined by a solid line with no reference to the applied data fitting procedure [17], which adds to the possible causes for the uneven shape of the diffraction pattern's envelope. Additionally, incompatibilities are detected between the diffraction pattern characteristics and the experimental setup (distance of double-slit from camera and possible deviation of the slit shape from rectangular).

The main measure that assesses the working hypothesis in Radin et al, the R parameter, is based on this problematic diffraction pattern from which the fast Fourier transform (FFT) power spectrum is extracted. The reported power spectrum, fig. 2B, of the double-slit diffraction intensity distribution, fig. 2A, do not seem to be compatible with each other. For instance, the spectrum appears to be erroneously positioned on the graph of fig. 2B. Also, efforts to estimate the fringe visibility, V, from the reported power spectrum as well as from the diffraction pattern could not yield compatible estimates, see appendix II.

Possible technical flaws and shortcomings in the experimental setup prepared to question the hypothesis in Radin et al could render the data on which their hypothesis is based as unreliable. Leaving for the moment this matter aside, it is primary to investigate whether the hypothesis itself is indeed testable under the proposed experimental setup. To be reminded, it was proposed that conscious attention distinguishes the path of a number of (indistinguishable) photons sent to a double-slit, instigating the collapse of their wavefunction. The collapse is manifested by the reduction of the visibility of the fringes inside the principal maximum of the diffraction pattern recorded by a digital camera.

Interference phenomena involving massive particles, e.g. atoms [8] or neutrons [4], can be explained only in quantum theory. On the other hand, experiments set to observe similar which-path quantum phenomena using photons would demand single photon interference experimental setups [6]. It has been emphasized that even if a strongly attenuated coherent laser beam is used, it would not be sufficient to illustrate quantum (particle) properties, such as the collapse of the wavefunction. Researchers are cautioned that the particular quantum state of the laser light in such cases is perfectly described as a wave by classical electrodynamics. Thus, it can never be used for which-path information measurements [10].

Attenuated laser beams in a double-slit diffraction experiment testing the interferometric complementarity (2) have been used without the need to introduce single photons, however, under an extra and important prerequisite; that the HeNe laser photons are simultaneously prepared in the same wavefunction produced by a single-mode beam [18]. The photons in the laser beam used by Radin et al do not fulfil this condition, since the beam operates in about three modes instead of one, according to the data in the manufacturer's manual (longitudinal mode spacing, 438 MHz: Melles Griot red HeNe laser 25 LHP 151). Also, the coherence length of the HeNe laser beam used in ref. [11] is approximately the same or less than the cavity length [19], which is about 34 cm and not more than one meter as it was suggested in ref. [11].

Considering the width of the HeNe laser beam (0.8mm according to the manufacturer's manual), the width of the slit (10μm) and its attenuated power (0.5mW), there should be about 10,000,000,000,000 photons on average of approximate energy $3 \cdot 10^{-19}$ J passing every second through each slit at the speed of light, on which the conscious attention of an observer is focused during the experiment. How many of this enormous number of indistinguishable photons can be "observed" during their flight from the laser to the camera that lasts about $0,0000000001$ seconds? The mechanism would seem incomprehensible by everyday experience, unless there is a process of interaction between consciousness and physical environment not yet scientifically described.

There is another reason why the hypothesis for a quantum process underway in Radin et al cannot be confirmed by the applied experimental setup. Collecting which-path-information on single photons or measuring the distinguishability parameter (P) is typically performed by use of physical apparatuses. This way the parameter P is quantified independently from the visibility parameter V. In the experiment of Radin et al. the photon distinguishability is not measured. Consciousness, or conscious attention, does not compare to the typical physical device that records photon trajectories. Unless there is an acceptable scientific theory to support that conscious attention is indeed affecting the photon distinguishability, any decrease of fringe visibility does not necessarily indicate that there has been an increase in photon distinguishability.

There can be a number of experimental factors responsible for the reduction of fringe visibility observed by a reliable experimental setup. For instance, the presence of more than one mode in the beam can create intensity instability and affect the coherence of the beam. Changes in camera sensitivity may be another factor, or even the presence of possible mechanical movements of the slits that may have caused the asymmetric diffraction pattern observed in fig. 2A. All things considered, until a scientific theory of interaction between consciousness and quantum particles (causing the collapse of the quantum wavefunction) is scientifically confirmed, the phenomena observed by attenuated laser beams should be fully interpreted by classical physics [20]. A fascinating concept such as the interaction of consciousness and matter would require the design of meticulous and carefully chosen experimental conditions. Till then, a safe distance from dangerous associations with quantum phenomena would be advisable.

In the words of Richard Feynman [21]:
*If science is to progress, what we need is the ability to experiment, honesty in reporting results – the results must be reported without somebody saying what they would like the results to have been – and finally – an*

*important thing – the intelligence to interpret the results.*

**Appendix I. The experimental set up of Radin et al.**

The intensity of diffracted light in such experiments consists of a succession of fringes, due to the interference of the light passing through each of the two slits. The fringes intensity, being maximum at the center, undulates (due to light diffracted through one slit) forming a central wide lobe, the principal maximum, and a series of narrower secondary maxima on either of its sides. At the ends of the central lobe the light intensity has diminished to zero marking the location of an absent fringe, the missing-order fringe, fig. 2.

(a) *The missing order fringe and the number of fringes inside the principal maximum.* The missing order fringe, $m_{missing\ order}$, is determined as [2, 22]

$$m_{missing\ order} = d/a = 20 \quad (I-1)$$

Therefore, the 20$^{th}$ order fringe inside the principal maximum should be missing, starting to count fringes from near the center where the zero-order fringe is located, fig.1. There should appear only 19 fringes on the left side and another 19 on the right side of the zero-order fringe (m=0), making the total number, N, of fringes to be

$$N = 2 \cdot (m_{missing\ order} - 1) + 1 = 39 \text{ fringes} \quad (I-2)$$

The missing order fringe in the diffraction pattern of ref. [11] is identified as the first fringe of minimum peak intensity. The apparent asymmetry in the diffraction pattern may be due to a slight angle of the slits relative to vertical, in the plane perpendicular to the incident beam [23], or a contamination of the pixels' surface. The position of diffraction fringes in fig.2A of ref [11] is given in pixels. The computer records intensities at each pixel between pixels 600 and 2400. These positions are slightly off the centers of the left and the right 13$^{th}$ fringe, respectively. They need to be shifted by about 19 pixels to the right along the x-axis so that the 13$^{th}$ fringe on either side of the centre falls exactly at the peak of the 600$^{th}$ and 2400$^{th}$ pixel, achieving also to bring the central fringe at the 1500$^{th}$ pixel, as it should be. Consequently, the order of missing fringe as it appears on this graph of fig. 2A in ref. [11] is the 16$^{th}$ on the right hand side and the 17$^{th}$ on the left hand side rather than the expected 20$^{th}$. This makes a total of 15+16+1=32 fringes observed inside the principal maximum of the reported diffraction pattern, less by 7 from the expected 39 fringes.

(b) *The width of the principal maximum.* The width of the principal maximum at the plane of the camera, the distance 2X between the centers of the two missing order fringes, is another feature that can be estimated in two ways. First, from the graph of fig. 2A of ref. [11] and also theoretically on the basis of the reported apparatus description. The peaks of the 16$^{th}$ on the right and the 17$^{th}$ fringe on the left are located at about 2623$^{th}$ and 315$^{th}$ pixel, respectively. The width of the principal maximum, as shown in fig. 2A of ref. [11] is, therefore, 2308 pixels.

Considering that the pixel width is $7 \cdot 10^{-6}$ μm, the width of the principal diffraction maximum at the plane of the camera is about 1.62 cm.

The expected width of the principal maximum, on the basis of the reported apparatus data is estimated as follows. The camera, fig.1, located at a distance 10.4 cm from the double slit, records the diffraction pattern along the line AB. The angle θ at which the slits view half of the central diffraction pattern on the plane where the camera is located (from the centre of fringe of order m=0 up to the missing fringe of order m=20 on either side of the centre, as described above), is determined as $\sin θ = λ/a$, which equals 0.06328 since $λ = 632.8$ nm and $a = 10$ μm. Therefore, $θ = \sin^{-1} 0.06328 \cong 0.0633$ rad, or 3.63°.

The distance X (the half width of the principal maximum) between the zero-order and the missing-order (m=20) fringe should, therefore, be $X = (10.4\ cm) \cdot \tan θ$ equal to about 0.659 cm. Consequently, the width 2X of the principal maximum is about 1.319 cm, or 1,884 pixels; narrower than the width of the principal maximum in fig. 2A of ref. [11] by 424 pixels.

(c) *The fringe separation at the plane of the camera.* The spacing between fringes, w, at the plane of the camera will be estimated both on the basis of the reported description of the apparatus and also from data of fig. 2A of ref. [11]. First, for distance D=10.4 cm and distance between slits $d = 2 \cdot 10^{-4}$ m [22]

$$w = D\frac{λ}{d} = 10.4 \cdot 10^{-2} m \cdot \frac{6.328 \cdot 10^{-7} m}{2 \cdot 10^{-4} m} = 3.29056 \cdot 10^{-4} m$$

(I-3)

This is equivalent to $(3.29056 \cdot 10^{-4} / 7 \cdot 10^{-6} \cong)$ 47 pixels.

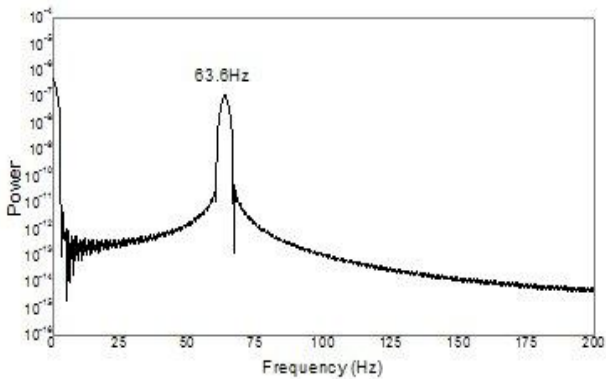

Fig. 3. Power spectrum of the simulated diffraction pattern of fig. 2.

The spacing between fringes, as it is illustrated on the diffraction pattern of fig. 2A of ref. [11], appears to depend on location. Between $600^{th}$ and $2400^{th}$ pixel (the peaks of the two fringes of $13^{th}$ order) there are 26 fringe separations. This makes about 69 pixels fringe separation as reported in the caption of fig. 2B of ref. [11]. If we assume the observed missing-order fringes, the $17^{th}$ on the left and the $16^{th}$ on the right, their distance is 2308 pixels spanning 33 fringe separations and yielding a fringe separation about 70 pixels instead; larger than what is expected on the basis of the apparatus description (47 pixels).

(d) *The position of the first secondary maximum.* The graph in fig. 2 shows a part of the simulated intensity distribution in the double-slit diffraction pattern, where the order of the missing fringe was chosen to be 20. The inset graph shows the whole pattern up to the $2^{nd}$ order secondary maxima. The intensity distribution $I(\theta)$ was plotted by use of the equation $I(\theta) = F(\theta) \cdot T(\theta)$, as the product of the diffraction term through one slit, $F(\theta)$ and the term of interference of light passing through each of the two slits, $T(\theta)$, where

$$F(\theta) = \left[\frac{\sin(10 \cdot \theta)}{(10 \cdot \theta)}\right]^2 \text{ and } T(\theta) = [\cos(200 \cdot \theta)]^2 \quad \text{(I-4)}$$

The simulation was performed by an Origin Pro® 7.5 custom program using 10004 data points of step $2 \cdot 10^{-4}$ (x-axis). The data points in this simulation are very narrowly spaced to ensure that there will be no fringe unobserved. The fringe distance in this simulation is 0.016 rad. Its associated peak in the FFT power spectrum is expected at $1/0.016 \cong 63$ Hz.

The peak of the $1^{st}$ secondary maximum in fig. 2A of ref. [11] is expected to be well within the observation field of the line camera. It should be located at about the $2869^{th}$ pixel. Its angular distance $\theta_2$ from the centre of the diffraction pattern, fig. 2, is determined by: $\sin\theta_2 = 1.45\lambda/a$, which equals 0.09176 [22]; this is equivalent to, $X_2$, where $X_2 = \tan\theta_2 \cdot D = 0.0922 \cdot 10.4 \text{ cm} \cong 0.958 \text{ cm}$ or, $X_2 = 0.958/7 \cdot 10^{-4} = 1369$ pixels away from the center. It is therefore located some $(1500 - 1369 =) 131$ pixels before the end of vision of the line camera. Its expected peak height is 0.047 times the peak intensity $I_0$ of the principal maximum (where:

**Table 1.** Fast Fourier transform frequencies in power spectrum

| Fourier frequency | 0 | 1 | K-1 | K | K+1 |
|---|---|---|---|---|---|
| Power | $\left(\frac{I_0}{2} + I_{DC}\right)^2$ | $\left(\frac{I_0}{2}\right)^2$ | $\left(\frac{I_0 V}{4}\right)^2$ | $\left(\frac{I_0 V}{2}\right)^2$ | $\left(\frac{I_0 V}{4}\right)^2$ |

$I_0 \cong 787$ a. u.) and its Y-coordinate about 282 arbitrary units; about twice the currently displayed intensity.

The observed width of principal maximum and spacing between fringes, both of them larger than expected, as well as the absence of the $1^{st}$ secondary maxima in fig. 2 of ref. [11] may all be due to a larger actual distance ($D_x$) of the camera from the slits than the reported 10.4 cm. If $D_{x_1}$ is its corrected value, considering the observed principal maximum width and $D_{x_2}$ its corrected value due to the observed fringe separation, then:

$$D_{x_1} = \frac{(0.81) \cdot (10.4)}{0.659} cm \cong 13 \ cm, D_{x_2} =$$

$$\frac{(69) \cdot 7 \cdot 10^{-6} m \cdot 2 \cdot 10^{-4} m}{6.328 \cdot 10^{-7} m} \cong 15 \ cm, \text{ where 0.81 cm is the}$$

observed half width of the principal maximum, yielding an average $\overline{D}_x \cong 14 \ cm$. If the camera is actually placed farther away from the double slit than its reported distance (10.4 cm), then the peak of the 1st secondary maximum would be well outside the view of the line camera and would therefore not be visible, as actually shown in fig. 2A of ref. [11].

(e) *The position of the characteristic peak in the FFT power spectrum.* Fig. 3 shows the fast Fourier transform (FFT) spectrum of the simulated data of fig. 2, performed by Origin Pro® 7.5. As expected, the spectrum exhibits a peak located at about 63 wavenumbers (the computer software assumes that the periodicity appears in time domain measured in seconds so that the frequency is measured in Hz), the inverse of the fringe spacing 0.016, (or more accurately the inverse of 0.0157).

The conversion of the peak position in wavenumbers associated with the fringe separation expressed in pixels goes as follows. To set the wavenumber unit of measurement, that is to convert from pixel$^{-1}$ to wavenumbers, the inverse of the sample length of 3000 pixels is being used: 1 pixel$^{-1}$ = 3000 wavenumbers. The inverse of the fringe spacing should appear on the power spectrum of the FFT transform as a peak at $(1/69)$ pixel$^{-1}$ = $(3000/69)$ wavenumbers $\cong 44$ wavenumbers. This is not the value indicated in the caption of fig. 2B of ref. [11] that is, the 45 wavenumbers. It is about one wavenumber less instead, close to where the power spectrum should be after shifting it by 1 wavenumber to the left. Simple inspection of the power spectrum in fig. 2B of ref. [11] shows that the spectrum is actually placed at the wrong position (in spite of what the figure caption reads) and must be shifted by about 1 wavenumber to the left until it touches the y-axis at zero. By shifting the graph, the position of the FFT peak becomes 44 wavenumbers, as it should be. As it will be shown later, this shift of spectrum to the left is required for the proper estimation of the R measure, according to the experimental analysis applied.

**Appendix II. Estimating the measure R in Radin et al.**

Consider a set of two slits of rectangular shape each of width, a, and separation, d, between their centers. The signal, $I(\theta)$, recorded by the camera at an angle $\theta$ from the center of the double-slit, fig. 1, is for uniformly illuminated slits:

$$I(\theta) = \frac{\sin^2(\pi a\theta/\lambda)}{(\pi a\theta/\lambda)^2} \cdot \cos^2\left(\frac{\pi d \sin\theta}{\lambda}\right) \approx$$

$$\approx F(\theta) \cdot \cos^2\left(\frac{\pi \cdot d \cdot \theta}{\lambda}\right), \text{ for small } \theta \quad \text{(II-1)}$$

where $F(\theta)$ represents the diffraction term of light passing through one slit.

The intensity distribution $I(\theta)$ can be approximated as [2]

$$I(\theta) = F(\theta) \cdot I_0[1 + V\cos(K \cdot \theta)] + I_{DC} \quad \text{(II-2)}$$

where the intensities of light at each slit are $I_1$ and $I_2$ and $I_0 = I_1 + I_2$. The arbitrarily added term $I_{DC}$ allows for the detector offset (about 245 arbitrary units, in fig. 2A of ref. [11]). The interference frequency K is expressed in wavenumbers.

The fringe visibility V in equation (II-2) is also defined as [2]

$$V = \frac{2\sqrt{I_1 I_2}}{I_1 + I_2}|\gamma_{12}(\tau)| \quad \text{(II-3)}$$

where $|\gamma_{12}(\tau)|$ represents the complex degree of coherence and $\tau$ is the time delay due to path difference in the propagation of the two beams. The intensity of the diffraction term across the camera width expands in a Fourier series

$$F(\theta) = a_0 + a_1\cos\theta + a_2\cos 2\theta + ... \quad \text{(II-4)}$$

and can be written in the following approximation

$$F(\theta) = \frac{1 + \cos\theta}{2} \quad \text{(II-5)}$$

normalized such that the maximum at $\theta=0$ is unity and the function goes to zero at the edges, so that

$$I(\theta) = \left(\frac{I_0}{2} + I_{DC}\right) + \frac{I_0\cos(\theta)}{2} + \frac{I_0 V\cos(K \cdot \theta)}{2} +$$
$$+ \frac{I_0 V\cos(K \cdot \theta) \cdot \cos(\theta)}{2} =$$
$$= \left(\frac{I_0}{2} + I_{DC}\right) + \frac{I_0\cos(\theta)}{2} + \frac{I_0 V\cos(K \cdot \theta)}{2} +$$
$$+ \frac{I_0 V\cos[(K+1) \cdot \theta]}{4} + \frac{I_0 V\cos[(K-1) \cdot \theta]}{4} \quad \text{(II-6)}$$

The correspondence between frequencies in the Fourier analysis and its power spectrum are shown in table 1. The peak in fig. 2B of ref. [11] is observed at K=45 wavenumbers. The frequencies at K-1 and K+1 represent two power spectrum side lobes at a lower strengths (by ¼). As K is not an integer, these adjacent peaks are not symmetrically located about K in the power spectrum.

The measure R can be estimated from the K-component (termed $P_D$ in Radin et al) of the spectrum with reference to the component at 1 wavenumber (termed $P_S$), where $P_S \equiv P_1$

$$R = \frac{D}{S} = \frac{\dfrac{P_D}{P_D + P_S}}{\dfrac{P_S}{P_D + P_S}} = \frac{P_K}{P_1} = \frac{(I_0 V/2)^2}{(I_0/2)^2} = V^2 \quad \text{(II-7)}$$

The relation (II-7) associates the R measure in Radin et al estimated from fig. 2B with the fringe visibility, V, estimated from fig. 2A.

After shifting the spectrum by one wavenumber to the left, as explained above, the strength of the power peak at now 44 wavenumbers measures about $10^{7.4}$, while the strength of the power at 1 wavenumber is about $10^{7.9}$. As a consequence of (II-7), since $R = P_K/P_1 \cong 0.32$, the visibility parameter will be $V \cong 0.56$.

In a similar fashion to the estimation of (II-7) a significantly better fit to the diffraction curve of fig. 2A in Radin et al is obtained by adding the term $A \cdot [\cos(2\theta - 1)]/2$ to $F(\theta)$ in (II-5) and assessing the parameter A. Now frequencies at wavenumbers 2, K-2 and K+2 will be added to those of table 1. A comparison between the new $F(\theta)$ and that in (II-4) (keeping only the three terms of the series) yields: $2a_0 = 1 - A$, $2a_1 = 1$ and $2a_2 = A$. A best fit is then obtained for A=0.25 ( $P_K/P_1 = 0.32$ ), which transforms (II-7) to

$$R = \frac{P_K}{P_1} = V^2\left(\frac{a_0}{a_1}\right)^2 \quad \text{(II-8)}$$

yielding $V \cong 0{,}75$.

The visibility parameter, V, can be directly estimated from fig. 2A. The intensity of fringes at the centre (at the zero-order fringe) gives $I_{max} \cong 1045$ a.u. and $I_{min} \cong 270$ a.u.. Considering the elevation $I_{elev} \cong 262$ a.u. of the diffraction pattern at the centre, the fringe visibility is

$$V^{(0)} = \frac{(I_{max} - I_{elev}) - (I_{min} - I_{elev})}{(I_{max} - I_{elev}) + (I_{min} - I_{elev})} \cong 0{,}98 \quad \text{(II-9)}$$

where $V^{(0)}$ denotes the visibility at the centre (zero order fringe). The visibility at the ends of the diffraction pattern is close to 100%, given that the minimum fringe intensity, $I_{min} - I_{elev}$, is practically zero. If the power spectrum were considered at its currently displayed position in fig. 2B of Radin et al, the power at 1 ($P_1$) —where the spectrum is very steep—would have been even higher than the observed $10^{7,9}$ units making the visibility even lower than 0.75. The above analysis renders the reported two graphs in figs. 2A and 2B in ref. [11] as incompatible.

**Acknowledgements** This work has benefited from the most valuable comments and suggestions of Dr Dave Andrews of the School of Physics & Astronomy at the University of Manchester, UK, to whom I am grateful. The Special Account for Research Grants of the University of Athens has financially supported this work.

**References**

1. G. Jaeger, A. Shimony, L. Vaidman Phys. Rev. A 51, 54 (1995).
2. E. Hecht, *Optics*, 4th ed, (Addison-Wesley Pub. Co., Reading, Mass., 2002).
3. W. K. Wootters, W. H. Zurek, Phys. Rev. D 19, 473 (1979).
4. D. M. Greenberger, A. Yasin, Phys. Lett. A 128, 391 (1988).
5. B.-G. Englert, Phys. Rev. Lett. 77, 2154 (1996).
6. P. D. D. Schwindt, P. G. Kwiat, B.-G. Englert, Phys. Rev. A 60, 4285 (1999).
7. V. Jacques, E Wu, F. Grosshans, F. Treussart, P. Grangier, A. Aspect, J-F Roch, Phys. Rev. Lett. 100, 220402 (2008).
8. S. Dürr, T. Nonn, G. Rempe Phys. Rev. Lett. 81, 5705 (1998).
9. P. Grangier, G. Roger, A. Aspect, Europhys. Lett., 1, 173 (1986).
10. C. Braig, P. Zarda, C. Kurtsiefer, H. Weinfurter, Appl. Phys. B 76, 113 (2003).
11. D. Radin, L. Mitchel, K. Galdamez, P. Wendland, R. Rickenbach and A. Delorme, Physics Essays 25, 2 (2012).
12. USB2.0 CCD Line Camera with External Trigger- LC1 USB at http://www.thorlabs.de/thorproduct.cfm?partnumber=LC1-USB
13. D. Slepian, JOSA 55, 1110 (1965).
14. G G Siut, L Chengt, D S Chius and K S Chanf, J. Phys. D: Appl. Phys. 27, 459 (1994).
15. X. Wu, Y. Wang, S. Li, C. Liu, EURASIP Journal on Advances in Signal Processing, 112 (2012).
16. L.T. Wood, arXiv:1207.4820v1 [physics.optics] (2012).
17. See for instance fig. 4 in ref.[23], fig.3 in ref. [24], or fig. 1 in ref. [25].
18. R Mir, J. S. Lundeen, M. W. Mitchell, A. M. Steinberg, J. L. Garretson, H. M. Wiseman, New Journal of Physics 9, 287 (2007).
19. R. S. Sirohi, *Course of Experiments with He-Ne Laser*, 2nd edition, (New Age International (P) Ltd Publishers, New Delhi, 1991), p. 14.
20. P. Mittelstaedt, A. Prieur, R. Schieder, Found. Phys. 17, 891 (1987).
21. Richard Feynman, The character of Physical Law. Probability and uncertainty. (The MIT Press, Cambridge, MA, 1985), at the end of chapter 6.
22. Jenkins FA and White HE, *Fundamentals of Optics,* p. 341 (McGraw Hill, New York, 1967) and Nelkon M. and Parker P. *Advanced Level Physics,* p. 491 (Heinemann Educational Books, London, 1982).
23. K K Gan and A T Law, Eur. J. Phys. 30, 1271 (2009). See page 1273.
24. W. Tittel, J. Brendel, B. Gisin, T. Herzog, H. Zbinden, N. Gisin, Phys. Rev. A 57 3229 (1998).
25. R. Tumulka, A. Viale, N. Zanghì, Phys. Rev. A 75, 055602 (2007) [4 pages].